\documentclass{iau} 
\usepackage{graphicx}

\title[UV-IR color profiles of 2K nearby galaxies] 
{UV-IR color profiles of the outer\\regions of 2K nearby S$^{4}$G galaxies}

\author[Alexandre Y. K. Bouquin \& Armando Gil de Paz]   
{Alexandre Y. K. Bouquin$^1$
 \and Armando Gil de Paz$^1$}

\affiliation{$^1$Departamento de Astrof\'{i}sica y CC. de la Atm\'{o}sfera, \\
Universidad Complutense de Madrid, \\
E-28040 Madrid, Spain \\
email: {\tt abouquin@fis.ucm.es} \\[\affilskip]}

\pubyear{2016}
\volume{321}  
\jname{Formation and evolution of galaxy outskirts}
\editors{A.C. Editor, B.D. Editor \& C.E. Editor, eds.}
\begin{document}

\maketitle

\begin{abstract}
We present our new, spatially-resolved, photometry in FUV and NUV from images obtained by GALEX, and IRAC1 (3.6 $\mu$m) photometry obtained by the Spitzer Space Telescope. We analyzed the surface brightness profiles $\mu_{\rm{FUV}}$, $\mu_{\rm{NUV}}$, $\mu_{[3.6]}$, as well as the radial evolution of the (FUV-NUV), (FUV - [3.6]), and (NUV - [3.6]) colors in the Spitzer Survey of Stellar Structures in Galaxies (S$^{4}$G) galaxies (d$<$40 Mpc) sample. We defined the GALEX Blue Sequence (GBS) and GALEX Red Sequence (GBR) from the (FUV - NUV) versus (NUV - [3.6]) color-color diagram, populated by late-type star forming galaxies and quiescent early-type galaxies respectively. While most disk becomes radially bluer for GBS galaxies, and stay constant for GRS galaxies, a large fraction ($>$50\%) of intermediary GALEX Green Valley (GGV) galaxies' outer disks are becoming redder. An outside-in quenching mechanism such as environmentally-related mechanisms such as starvation or ram-pressure-stripping could explain our results.
\end{abstract}

\firstsection 
              
\section{Introduction}

%
It has been seen, in several nearby disk galaxies, that SF can occur 
at least as far as 4 or 5 times their optical radius D25,
where average gas densities are very low.
This means that there are locally dense regions of gas in the outskirts capable
of collapsing into stars, but the mechanism that triggers such collapse is yet to be known.
One possibility would be the infall of gas from an external reservoir of gas
surrounding galaxies (e.g. \cite[S{\'a}nchez Almeida et al. 2014]{SanchezAlmeida2014}).
Evidence of the existence of such reservoir can be seen in
the HI gas distribution around galaxies (e.g. \cite[Simkin et al. 1987]{Simkin1987}, \cite[Tilanus \& Allen 1993]{Tilanus1993}),
extending further out than the UV-disks, and could well be the source for such cosmic accretion (\cite[Sancisi et al. 2008]{Sancisi2008}).
These would support the \textit{inside-out} disk growth scenario.

In this article, we briefly explain the sample we used and the analysis we performed.
We discuss the implications of our analysis in the last section.
Throughout this article, magnitudes are in the AB system, and the Hubble constant is 75 km/s/Mpc.

\section{Sample and Analysis}

Our base sample is the \textit{Spitzer Survey of Stellar Structure in Galaxies} (S$^{4}$G) \cite[Sheth et al. 2010]{Sheth2010}
which consists of 2300 nearby galaxies within 40 Mpc at galactic latitudes $|b| > 30^{\circ}$,
and B-band apparent (Vega) magnitude corrected for extinction $m_{\rm{Bcorr}}$ (Vega) $<$ 15.5, with angular diameter $\theta > 1'$.
We also make extensive use of the S$^{4}$G IRAC1 3.6 $\mu$m photometry obtained by \cite[Mu\~{n}oz-Mateos et al. (2015)]{MunozMateos2015}.

We gathered GALEX GR6/7 FUV and NUV images,
performed a semi-automatic, semi-manual masking,
sky-subtraction, and obtained photometry as in \cite[Gil de Paz et al. (2007)]{gdp2007B}.
As shown in \cite[Bouquin et al. (2015)]{Bouquin2015}, the (FUV-NUV) vs (NUV-[3.6]) color-color diagram,
constructed from asymptotic magnitudes, revealed the clearly bimodal distribution of nearby galaxies,
with a narrow blue sequence (the GALEX blue sequence or GBS) populated by late-type star forming galaxies (SFG),
and a less narrow red sequence (the GALEX red sequence or GRS) populated by quiescent early-type galaxies distributed along,
while some intermediary type galaxies are distributed between the GBS and the GRS (the GALEX green valley or GGV), 
which are seemingly leaving the GBS and going toward the GRS (although the possibility for a galaxy to go from the GRS to the GBS is not ruled out).
While the GBS extends and intersects the GRS, galaxies in the GGV seems to be drifting away from the GBS , effectively becoming redder.
So what sets these GGV galaxies apart from other GBS or GRS galaxies? It is clearly not driven by mass since more massive GBS are found.
%

In order to evaluate the reddening/blueing of the outer disk quantitatively, 
we first identified the disk region in our galaxies using an homogeneous set of criteria.
From the 3.6 $\mu$m surface brightness (SB) versus kiloparsec profiles,
we fit an error-weighted line to the disk part once the region dominated by the disk light is delimitated.
We used the galactocentric distance R80 that corresponds to the distance where 80\% of the IR light is enclosed,
and varied our disk cutoff by factors of R80 to exclude inner data points to decide where the best linear fit occurs.
We also imposed a SB magnitude cutoff to select only the fainter parts of the disk.
We find that the best fit, with a close-to-unity average reduced $\chi^{2}$, occurs when R/R80=0.25 to 0.50 and $\mu_{[3.6]}$ = 22.5 to 23 mag/arcsec$^{2}$.

We obtained the color gradients of ($\mu_{\rm{FUV}} - \mu_{\rm{NUV}}$), 
($\mu_{\rm{FUV}} - \mu_{[3.6]}$), and ($\mu_{\rm{NUV}} - \mu_{[3.6]}$) colors versus galactocentric distances in kiloparsec.
These color profiles show that the ($\mu_{\rm{FUV}} - \mu_{\rm{NUV}}$) color is most sensitive to variations in the recent SF history:
while most galaxies have bluer outer disks, a significant fraction ($>$50\%) of GGV galaxies have a more positive gradient (i.e. reddening).
This is also seen in these other colors.

Besides the determination of the color gradients,
we performed the same exercise with the 3.6 $\mu$m SB profiles of disk-models of \cite[Boissier \& Prantzos (2000)]{BP00} (BP00) generated from various circular velocities $V_{\rm{c}}$ ranging from 20 to 430 km/s with a 10km/s step, and spin parameters $\lambda$ ranging from 0.002 to 0.150 with a 0.001 step.
The total number of models generated is 6258.
In this manner, we successfully obtained a pair of slope and y-intercept from each galaxy and each model,
which we show on the same plot (Fig.1 left), and could determine the $\lambda$ and $V_{\rm{c}}$ for a large and
well-defined sample of $\sim$2000 nearby galaxies.

\begin{figure}[b]
\begin{center}
\includegraphics[width=2.5in]{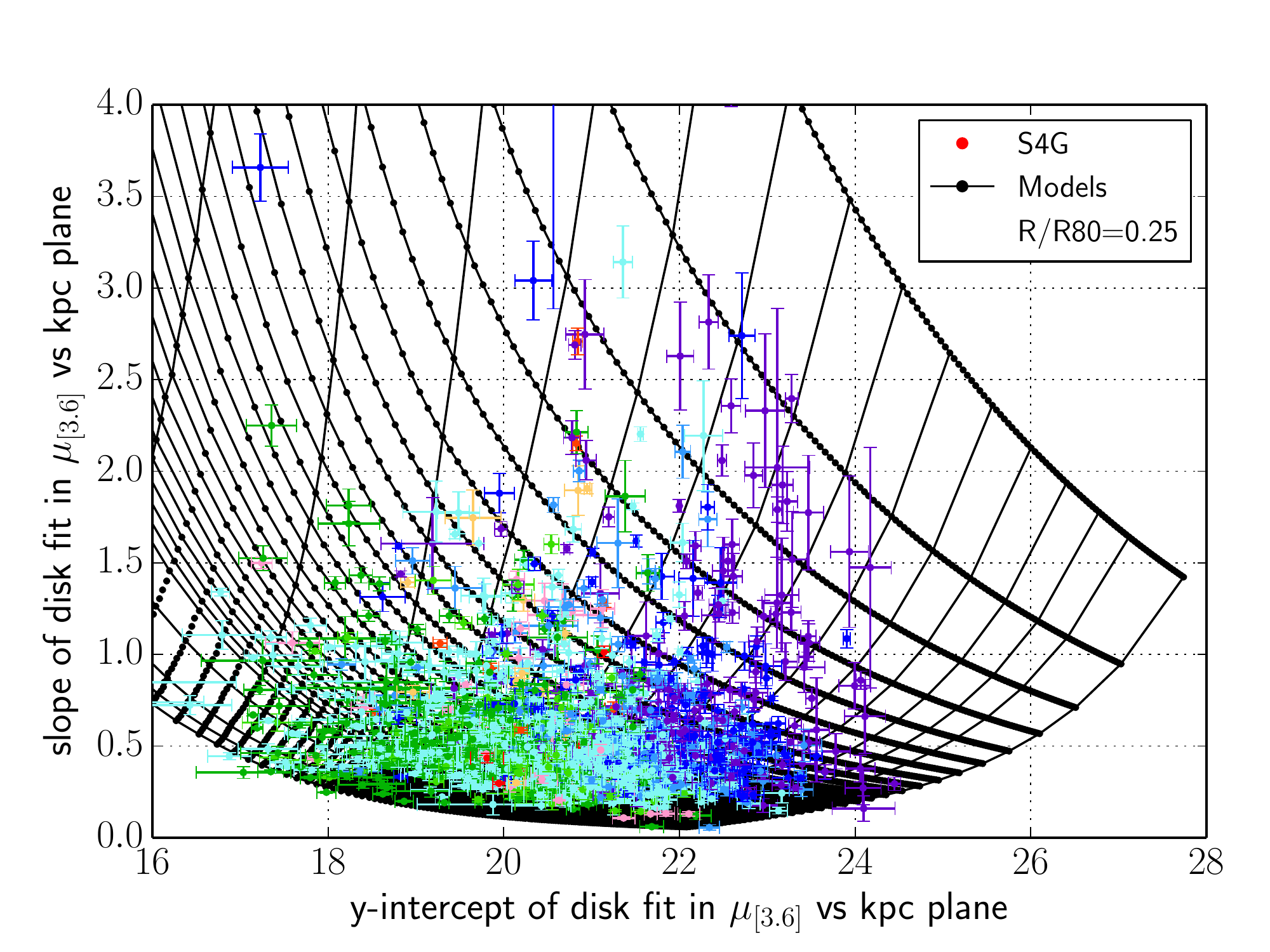}
\includegraphics[width=2.5in]{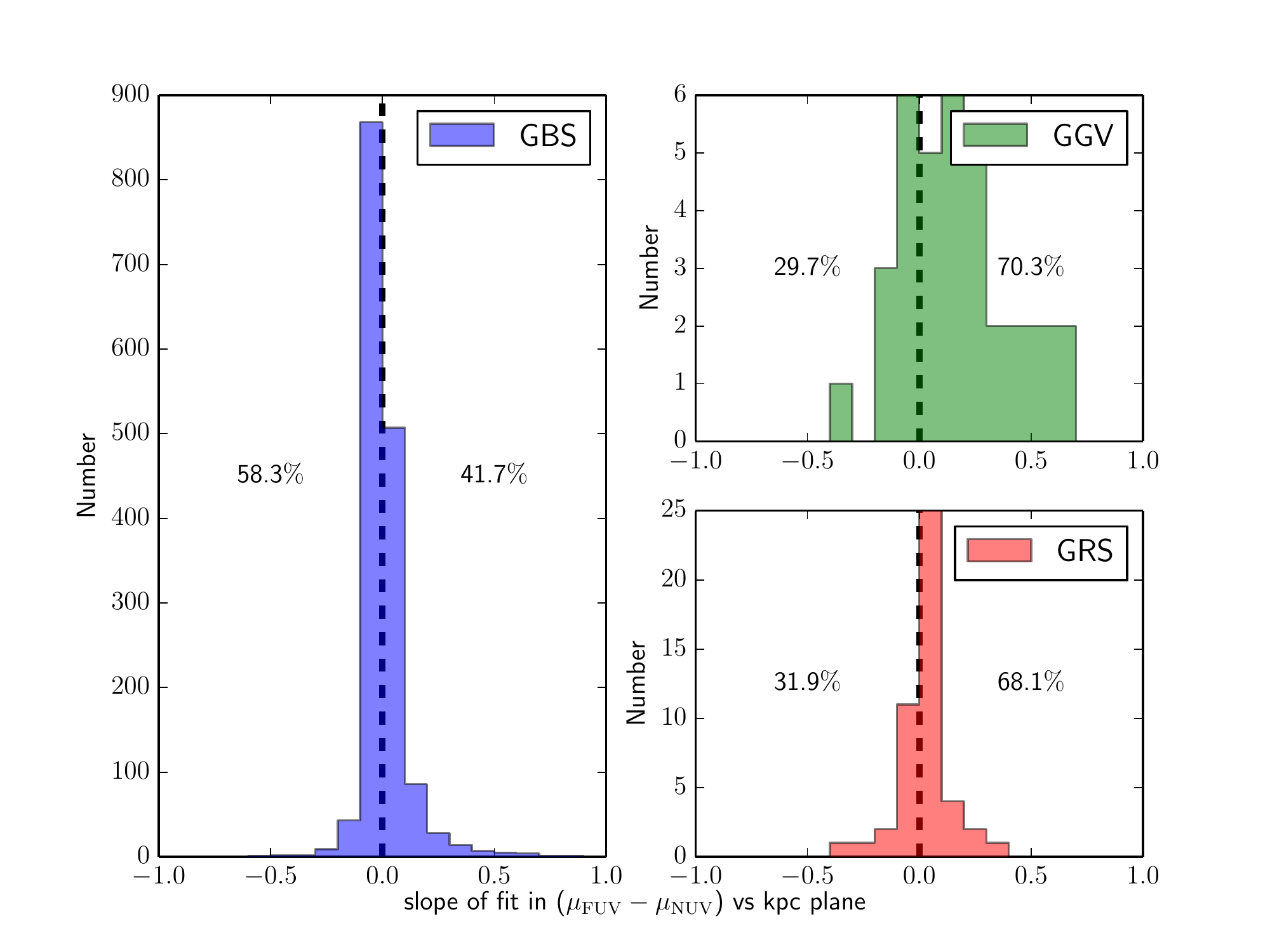} 
 \caption{\textit{Left}: slope vs y-intercept of BP00 models (black grid) of varying $V_{\rm{c}}$ (increasing from top to bottom, annotated V) and $\lambda$ (increasing from left to right, annotated L), and of our galaxy sample (colored points). The color-coding is taken from \cite[Bouquin et al. (2015)]{Bouquin2015} and corresponds to morphological types. The slopes and y-intercepts were obtained from an error-weighted linear fit to the $\mu_{[3.6]}$ SB profiles. 
 \textit{Right:} Histograms of the slope of the fits for GBS, GGV, and GRS galaxies for linear fits done in the ($\mu_{\rm{FUV}} - \mu_{\rm{NUV}}$) vs kpc plane.
The percentages shown are the fractions of negative slopes (left) and positive slopes (right) in each of the subsample.
Bin width is 0.1 in all cases. The histograms are constrained to slopes between -1 and 1 for clarity.
}
\label{fig1}
\end{center}
\end{figure}


\section{Conclusions}

We obtained FUV, NUV, IRAC1 photometry from the S$^{4}$G sample for over 2000 nearby galaxies.
We used the IRAC1 3.6$\mu$m SB profiles,
and fitted a line to their disk part, obtaining a set of slope and y-intercept for each galaxy.
We also used spatially-resolved ($\mu_{\rm{FUV}} - \mu_{\rm{NUV}}$), ($\mu_{\rm{FUV}} - \mu_{[3.6]}$), and ($\mu_{\rm{NUV}} - \mu_{[3.6]}$) colors to study their gradients.
We found that a higher fraction of GGV galaxies have positive gradients (i.e. reddening disks) compared to GBS and GRS galaxies in all these colors.
The gradient of ($\mu_{\rm{FUV}} - \mu_{[3.6]}$) color translates to a constant, albeit lower and redder, 
sSFR ($\log$(sSFR)$\sim$-12) for GRS and GGV galaxies while in general, the sSFR tends to increase as we move to the outer disks.

These reddening disks always have a redder outskirts than their inner part,
with a bluer bulge and red disk, and appear to be reddening \textit{outside-in}.
This indicates that whatever mechanism(s) is/are quenching SF activities
starts affecting the outskirts of disks first and eating its way to the inner parts.
This could be explained by starvation,  strangulation (\cite[Somerville et al. 2008]{Somerville2008}, \cite[Larson et al. 1980]{Larson1980}), 
mergers and interactions (\cite[Toomre \& Toomre 1972]{Toomre1972}),
or ram-pressure stripping (\cite[Gunn \& Gott 1972]{Gunn1972}),
effectively removing low-density gas first in the outskirts.

The same exercise was done over 6000 BP00 disk models, creating a grid of slopes and y-intercepts.
By relating the slopes and y-intercepts of the data to the models, we obtained $V_{\rm{c}}$ and $\lambda$ for each galaxy.
These will be the subject for further analysis.

We acknowledge financial support to the DAGAL network from the People Programme (Marie Curie Actions)
of the European Union’s Seventh Framework Programme FP7/2007-2013/ 
under REA grant agreement number PITN-GA-2011-289313, and
financial support from the Spanish MINECO under grant number AYA2013-41243-P.
A.G.d.P. acknowledges financial support under grant number AYA2012-30717.

\end{document}